\def\lesssim{\mathrel{\hbox{\rlap{\hbox{\lower4pt\hbox{$\sim$}}}\hbox{$<$}}}}

\def\gtrsim{\mathrel{\hbox{\rlap{\hbox{\lower4pt\hbox{$\sim$}}}\hbox{$>$}}}}

\def\msun{$M_{\odot}$}

\def\teff{$T_{\rm eff}$~}

\def\ll_lsun{log$({L/\rm L_{\odot}})$~}

\def\masa_msun{$M/ \rm M_{\odot}$~}

\def\m_mstar{$M/M_{*}$~}

\documentclass{aa}
\usepackage{graphicx}

\begin{document}

\title{DQ white-dwarf stars with low C abundance: Possible progenitors}

\author{C. G. Sc\'occola$^{1}$\thanks{Fellow of CONICET, Argentina}, 
       L. G. Althaus$^{1}$\thanks{Member of the Carrera del 
       Investigador Cient\'{\i}fico y Tecnol\'ogico, CONICET, Argentina.}, 
       A. M. Serenelli$^{2}$, R. D. Rohrmann$^{3\star\star}$ \and 
       A. H. C\'orsico$^{1\star\star}$}
\offprints{C. G. Sc\'occola}

\institute{$^1$Facultad de Ciencias Astron\'omicas y  Geof\'{\i}sicas, 
	   Universidad Nacional de La Plata, Paseo del Bosque, s/n,
	   (1900) La Plata, Argentina.\\ Instituto de Astrof\'{\i}sica
	   La Plata, IALP, CONICET.\\
	   $^2$Institute for Advanced Study, School of Natural
	   Sciences, Einstein Drive, Princeton, NJ, 08540, USA.\\
	   $^3$Observatorio Astron\'omico, Universidad Nacional de
	   C\'ordoba, Laprida 854, (5000) C\'ordoba, Argentina.
\email{cscoccola,althaus,acorsico@fcaglp.unlp.edu.ar, aldos@ias.edu, rohr@oac.uncor.edu}}
\date{Received; accepted}

\abstract{The present paper focuses on the evolution of
hydrogen-deficient white dwarfs with the aim of exploring the
consequences of different initial envelope structures on the carbon
abundances expected in helium-rich, carbon-contaminated DQ white
dwarfs. In particular, the evolutionary link between the DQs with low
detected carbon abundances and the PG1159, extreme horizontal branch,
and helium-rich R Coronae Borealis (RCrB) stars is explored. We
present full evolutionary calculations that take a self-consistent
treatment of element diffusion into account as well as expectations
for the outer layer chemical stratification of progenitor stars upon
entering the white dwarf regime. We find that PG1159 stars cannot be
related to any DQ white dwarfs with low C abundances. Instead, we
suggest that the latter could constitute the progeny of the giant,
helium-rich RCrB stars.
\keywords{stars:  evolution  --  stars: interiors -- stars:
white dwarfs -- stars: oscillations  } }  
             
%\abstract{}{The present paper focuses on the evolution of
%hydrogen-deficient white dwarfs with the aim of exploring the
%consequences of different initial envelope structures on the carbon
%abundances expected in helium-rich, carbon-contaminated DQ white
%dwarfs. In particular, the evolutionary link between the DQs with low
%detected carbon abundances and the PG1159, extreme horizontal branch,
%and helium-rich R Coronae Borealis (RCrB) stars is explored.}{We
%present full evolutionary calculations that take a self-consistent
%treatment of element diffusion into account as well as expectations
%for the outer layer chemical stratification of progenitor stars upon
%entering the white dwarf regime.}{We find that PG1159 stars cannot be
%related to any DQ white dwarfs with low C abundances. Instead, we
%suggest that the latter could constitute the progeny of the giant,
%helium-rich RCrB stars.}{}

%\keywords{stars:  evolution  --  stars: interiors -- stars:
%white dwarfs -- stars: abundances  }

\authorrunning{Sc\'occola et al.}

\titlerunning{DQ white dwarf progenitors}

\maketitle

%----------------------------------------------------------------

\section{Introduction}

White dwarfs constitute the final evolutionary stage for the vast
majority of stars.  Standard stellar-evolution theory predicts that
most white-dwarf progenitors are post-asymptotic giant branch (post-AGB)
stars of low and intermediate masses that enter the hot white-dwarf
stage with hydrogen-rich surface layers (DA spectral type). However,
about 20 \% of all observed white dwarfs are characterized by
hydrogen-deficient atmospheres (commonly referred to as DB spectral
type or more appropriately non-DA), and their existence has been a bone
of contention amongst researchers for many years.  The most commonly
accepted mechanism by which hydrogen-deficient white dwarfs are formed
 is one that involves the occurrence of a final thermal pulse during
the post-AGB evolution (the born again scenario; see Fujimoto 1977,
Sch\"onberner 1979, and Iben et al. 1983 for earlier references, and Bl\"ocker
2001 for a recent review).  In recent years, evolutionary
channels other than those involving post-AGB evolution that could
lead to the formation of hydrogen-deficient white dwarfs have been
suggested.  Amongst them, we find the hydrogen-poor, carbon-normal
star evolution directly from the extreme horizontal branch (AGB
manqu\'e stars such as sdB stars) into the white dwarf state and the
evolution of helium-rich supergiants R Coronae Borealis (RCrB) stars
(Sch\"onberner 1996 and De Marco et al. 2002 for a recent reference).
In particular, sdB stars are hot evolved stars with stellar masses
around 0.5 \msun\ that populate the extreme horizontal branch. Their
envelopes are so thin that the AGB phase of evolution after core
helium exhaustion is avoided. Instead, they  evolve directly into
low-mass white dwarfs (see Fontaine \& Chayer 2005 for a recent
summary of their properties).

 Within the born-again scenario, most of the residual hydrogen is
completely burnt (see Herwig et al. 1999 and Althaus et al.  2005 for
modern computations). The remnant evolves into the central star of a
planetary nebulae at a high effective temperature (\teff) as a
hydrogen-deficient, helium-burning object with carbon-rich outer
layers; see Althaus et al. (2005) for a complete simulation of the
formation and evolution of hydrogen-deficient white dwarfs through a
born-again episode.  Observational examples of these
hydrogen-deficient post-AGB stars are the very hot PG1159 and their
probable progenitors, the Wolf-Rayet type central stars of planetary
nebulae having spectral type [WC] (Koesterke \& Hamann 1997; Dreizler
\& Heber 1998; Werner 2001).  Spectroscopic
analyses reveal that most of these stars are characterized by surface
layers rich in helium, carbon, and oxygen with a typical composition
(in mass fraction) of (He, C, O)= (0.33, 0.50, 0.17) (Werner 2001).

The analysis by Dreizler \& Heber (1998) suggests that PG1159 stars
are the direct predecessors of the majority of helium-rich DO white
dwarfs, the hot and immediate progenitors of DB white
dwarfs. Theoretical calculations that incorporate the effect of
element diffusion and winds also predict the existence of a link
between the PG1159 stars and most of the DO's (Unglaub \& Bues 2000).
In this context, evolutionary calculations taking time-dependent
element diffusion into account (Dehner \& Kawaler 1995 and Gautschy \&
Althaus 2002) have shown that, as a result of the gravitational
settling of carbon and oxygen, PG1159 stars are expected to evolve
into white-dwarf stars with a superficially helium-dominated
double-layered chemical structure.  In fact, two different chemical
transition zones would characterize the envelope of the PG1159
descendants: a still uniform intershell region rich in helium, carbon,
and oxygen, which are the relics of the short-lived mixing episode
that occurred during the last helium thermal pulse, and an overlying
pure helium mantle that thickens as cooling proceeds.

 Another group of hydrogen-deficient stars are the extreme helium
stars (EHes), which are often considered to be related to the cooler
helium-rich RCrB stars because of their similar surface composition
and luminosities.  As recently discussed by Saio \& Jeffery (2002),
the EHes could be formed through a merger of a carbon-oxygen white
dwarf with a less massive helium white dwarf.

The development of a double-layered chemical structure in
hydrogen-deficient white dwarfs requires initial carbon to be abundant
throughout the outer layers of the star.  A viable route for this to
occur is that a final thermal pulse takes place during the post-AGB
evolution.  However, the existence of DB white dwarfs characterized by
a single-layered profile even from the very beginning of their
evolution cannot be discounted, if an evolutionary link between the
RCrB or the AGB manqu\'e stars and hot hydrogen-deficient white dwarfs
indeed exists.

From the above concerns, it seems quite reasonable to suppose that DB
white dwarfs could harbour envelopes with markedly distinct abundance
profiles.  This issue motivated us to undertake the present
investigation, which is essentially focused on DQ white dwarfs, the
supposed cooler descendants of DBs.  DQ white dwarfs have \teff\
values below 13000 K (Weidemann \& Koester 1995, see also Bergeron et
al. 2001) and are characterized by trace amounts of carbon in their
atmospheres.  Their abundance by number relative to helium, log
($n_{\rm C}/n_{\rm {He}}$), ranges from -7.3 to -1.5 (MacDonald et al.
1998).  The origin of carbon in the atmosphere of these stars is
thought to be the result of convective dredge-up of the carbon
diffusive tail by the helium convection zone (Pelletier et al.  1986;
see also Koester et al.  1982). DQ stars are particularly interesting
because the wide range of surface carbon abundances they exhibit could
be reflecting the existence of hydrogen-deficient white dwarfs with a
variety of evolutionary history and formation processes.  In
connection with this, MacDonald et al. (1998) have suggested that the
diversity of C to He ratios is due to a wide range of white-dwarf
masses. Indeed, under the assumption that DQs have predecessors with
single-layered chemical profiles, they propose that those with very
high C to He ratios are significantly more massive than the typical
white dwarf, say 1.0 $M_{\odot}$, and that they have envelope masses
running from 10$^{-4}$ to 10$^{-3}$ \msun. Furthermore, they speculate
that DQ with the lowest C to He ratios can be understood by allowing
the WD to be less massive than the canonical 0.6 \msun.

We reexamine the carbon abundances in DQs in the framework of
detailed evolutionary calculations. The present paper is not intended
to explain the whole range of observed C to He ratios, but instead we
focus on those DQs with low carbon abundances. This work is motivated
by the recent suggestion by Althaus \& C\'orsico (2004) that DQs with
{\it low} detected carbon would not fit into the evolutionary
connection PG1159$\rightarrow$ DB $\rightarrow$ DQ.  In this work, we
explore alternative evolutionary channels (with realistic initial
outer layer stratification) that could lead to DQs with low carbon
abundances. The study of the consequences of structurally different
initial envelopes for the carbon surface abundances in DQ constitutes
the main purpose of our work. It is worth mentioning that Dufour et
al. (2005) have presented new theoretical calculations aimed at
explaining the monotonic decrease in carbon pollution in cool DQs with
decreasing \teff.

In what follows we provide details about the physical ingredients
involved in the calculations, as well as the model sequences
studied. In Sect. 3 we describe the main results.  We conclude the
paper in Sect. 4 with a summary of the results and some final remarks.

\section{Input physics and evolutionary sequences}

The evolutionary code used in this work is the DB white-dwarf
evolutionary code recently employed in Gautschy \& Althaus (2002) and
Althaus \& C\'orsico (2004), so we refer the reader to those works for
details. The important point here is that this code allows us to
simulate the white-dwarf evolution in a self-consistent way including
abundance changes resulting from time-dependent element diffusion.
Our diffusion treatment, based on a formulation for multicomponent
gases by Burgers (1969), accounts for gravitational settling and
chemical and thermal diffusion for the nuclear species $^{4}$He,
$^{12}$C, and $^{16}$O. Diffusion velocities were evaluated at each
evolutionary time step. The abundance changes caused by diffusion were
treated separately from those resulting from convective mixing. The
microphysics included an updated version of the equation of state of
Magni \& Mazzitelli (1979), OPAL radiative opacities for arbitrary
metallicity (Iglesias \& Rogers 1996) including carbon- and
oxygen-rich compositions, up-to-date neutrino emission rates (Itoh
1997), and conductive opacities (Itoh et al 1994).  Convection is
treated in the framework of the mixing length theory as given by the
ML2 parametrization (see Tassoul et al. 1990).  As for expectations
for the carbon enrichment during the DQ stage, the treatment of
convection is hardly relevant, since the depth reached by
the superficial helium convection zone became almost insensitive to
the efficiency of convection by the time the domain of the DQs was
reached. We assumed complete ionization for
calculating diffusion velocities. This prevented us from being
confident about the predictions of our models for the carbon
enrichment in DQs with \teff $\lesssim$ 9000 K.

The initial stellar models needed to start our cooling sequences were
obtained by means of the same artificial procedure as described in
Gautschy \& Althaus (2002).  As one of our main purposes was to
understand what could be the predecessors of DQs with very low C
abundance, and because different evolutionary channels lead to
distinct chemical profiles, we classified the computed sequences
according to their initial chemical structure.  Hence, we assumed
the following types of initial hydrogen-deficient white dwarf
configurations:

\begin{itemize}

\item As for the first case, which will be hereinafter referred to as 
double-layered models, we assumed an initial uniform envelope chemical
composition  of helium, carbon,  and oxygen, which is representative of PG1159
stars, that  is, DB white dwarf  progenitors that have emerged from a
late  helium thermal  pulse  on  the early  cooling  track (Herwig  et
al. 1999).
 
\item As for the second case, referred to as single-layered models,
we assumed envelopes that are a mixture of helium and trace
carbon. There, helium is almost completely separated from carbon
even from the onset of the hot phases of the evolution.  This
chemical stratification is appropriate for hot hydrogen-poor white
dwarf progenitors such as the extreme helium stars and the RCrB
stars. It could also be appropriate for AGB manqu\'e stars. Note that
for these models we assumed $X_C=0.001$. This is suitable for the
carbon abundance detected in some RCrB, as recently listed in De Marco
et al. (2002) (see their Table 5). For our purposes, we did not
consider the presence of trace amounts of oxygen in the outer
layers.

\end{itemize}

For both types of sequences we considered models with stellar masses
of 0.50, 0.60 and 0.70 \msun\ and with helium envelope mass ($M_{\rm
He}$) in the range $0.001 \lesssim M_{\rm He}/M_* \lesssim 0.02 $. The
elected $M_{\rm He}$ range amply covers what is expected for the white
dwarf progenitor from stellar evolution theory. This parameter space
allows us to explore the dependence of the carbon enrichment at some
length.  The chemical composition of the core is from Salaris et
al. (1997).

Additionally, we computed the full evolution of a 0.49 \msun\
model from the extreme horizontal branch (EHB) down to the domain of
the DQ white dwarfs with the aim of covering the possible
evolutionary connection between DB white dwarfs and the AGB manqu\'e
stars. The outer chemical structure resembles that of
a single-layered configuration.

In Table 1 we list the whole set of model sequences to be discussed in
this paper.  We use a condensed notation to identify a specific
sequence.  For instance, 06DL1e3 stands for the double-layered
0.6-\msun\ sequence with a helium envelope of $0.001M_*$ (DL and SL
refer, respectively, to double- and single-layered models).  The 05EHB
corresponds to the sequence computed from the extreme horizontal
branch.  For each sequence, we list the stellar mass in solar units,
the mass fraction of helium content\footnote{ M$_{He}$ is the mass
fraction of helium content in the whole white dwarf. In the case of SL
models, He constitutes the envelope, while in the case of DL models,
He is distributed in two zones: the pure helium envelope, which
thickens as cooling proceeds, and the zone that lies between the
latter and the C-O core, which consists of a mixture of He, C, and O.}
in the star, and the initial abundance (by mass) of surface carbon.  In
all of these cases, evolution was computed self-consistently with
time-dependent element diffusion from high effective temperatures down
to below \teff 10000 K.

\begin{table}
\caption{Model sequences employed in our study. }
\renewcommand{\arraystretch}{1.3}
\begin{tabular}{p{1.5cm}p{1.5cm}p{1.5cm}p{1.5cm}}
\hline
Sequence &  $M_*$/\msun &  $M_{\rm He}/M_*$ & 
X$_{^{12}{\rm C}}^{\rm sur}$\\
\hline
 05DL1e2 & 0.50 & 0.01 & 0.36\\
 05DL2e2 & 0.50 & 0.02 & 0.36\\ 
 06DL1e2 & 0.60 & 0.01 & 0.36\\ 
 06DL1e3 & 0.60 & 0.001 & 0.36\\
 07DL1e2 & 0.70 & 0.01 & 0.36\\
 07DL1e3 & 0.70 & 0.001 & 0.36\\
 05SL1e2 & 0.50 & 0.01 & 0.001\\
 05SL2e2 & 0.50 &  0.02 & 0.001\\
 06SL1e2 & 0.60 &  0.01 & 0.001\\
 06SL1e3  & 0.60  & 0.001 &  0.001\\
 07SL1e2  & 0.70 &  0.01 & 0.001\\ 
 07SL1e3 & 0.70 & 0.001 & 0.001\\ 
 05EHB & 0.49 & 0.08 & 0.03 \\
\hline
\end{tabular}
\end{table}

In what follows, we describe the main results of our calculations. Our
main interest is concentrated on expectations for the carbon
enrichment in DQ white dwarfs. At such advanced stages of evolution,
the artificial way our models are generated is not expected to
significantly alter the results to be presented.  Hence, a detailed
treatment of the evolutionary phases prior to the formation of DB
white dwarfs is not of primary importance for our purposes.

\section{Evolutionary results}

\subsection{DL and SL sequences}

Throughout the present work we shall assume that convective dredge-up
is the mechanism responsible for the observed carbon abundances in DQ
white dwarfs.  In this section we shall investigate at some length the
role played by two issues that are intimately related to the occurrence of
convective dredge-up.  These are the outer convection zone
(hereinafter referred to as OCZ) and the shape of the outer chemical
profiles.  Specifically, what determines the amount of carbon that
eventually will be dredged to the stellar surface are (i) the location
of the bottom of the OCZ, and (ii) the presence of a tail of the
$^{12}$C distribution at the intershell region. We start our analysis
by examining the evolution of the OCZ for white dwarf models with
different stellar masses and helium contents, and a distinct chemical
structure of their envelopes. We then describe the evolution of the
chemical profiles and of the C to He ratios, comparing what happens in
different sequences.

\subsubsection{Outer convective zones}

\begin{figure}[ht]
\centering
\includegraphics[clip,width=250pt]{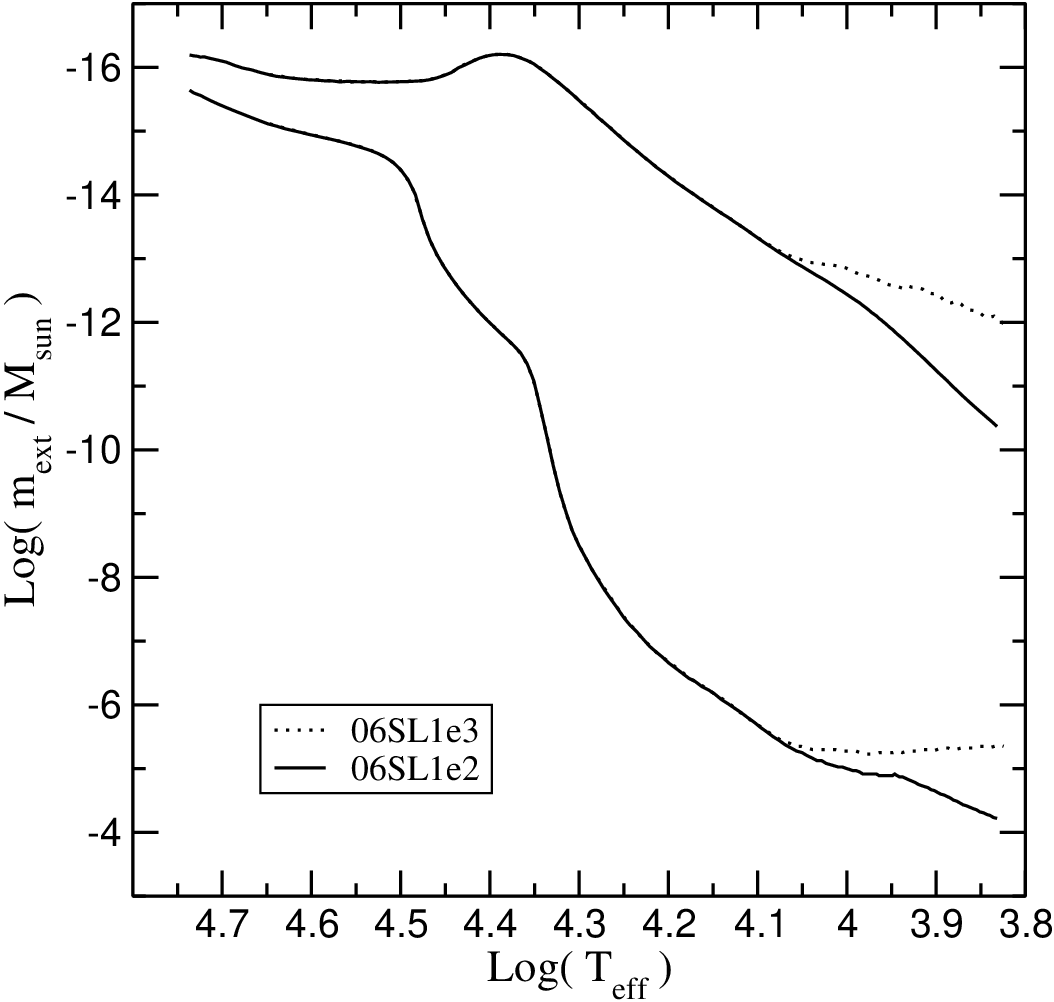}
\caption{ Top and bottom layers of the OCZ expressed in terms of $\log
(m_{\rm ext}/M_{\odot})$ versus $\log(T_{\rm eff})$ (being $m_{\rm
ext}= M_* - M_r$) for 0.6-$M_{\odot}$ single-layered models with
$M_{\rm He}= 10^{-2} M_*$ (solid line) and $M_{\rm He}= 10^{-3} M_*$
(dotted line). For the model with thinner helium envelope the bottom
of the OCZ remains shallower from $T_{\rm eff} \sim 11200$ K.}
\label{grafico1}
\end{figure}

In Fig. \ref{grafico1}, the boundaries of the OCZ as measured by the
outer mass fraction are depicted in terms of the effective temperature
for sequences 06SL1e3 and 06SL1e2. These two sequences only differ in
the mass of the helium layer.  The size of the OCZ is almost identical
for both sequences, except when the models reach low effective
temperatures, say $\log(T_{\rm eff}) \lesssim 4.1$.  From there, the
bottom of the OCZ for models with thin helium envelopes remains at a
fixed maximum depth, whereas the OCZ of the thick helium envelope
continues to grow farther inwards.  This is because at low effective
temperatures, the convective layers have reached the tail of the
carbon distribution and consequently some carbon has begun to be
dredged to the surface, causing a rise in the opacity throughout the
outer layers. As a result, the temperature gradient increases and the
region of partial ionization --- which is coincident with the location
of the OCZ --- rises.  This effect, coupled with the fact that the
white dwarf is cooling, prevents the OCZ from growing more.  On the
contrary, the white dwarf models with thick helium envelopes does not
suffer from this incipient contamination of carbon, and as the white
dwarf cools, the OCZ reaches deeper layers.

Now, we explore the effect of varying the stellar mass, holding the
thickness of the helium layer fixed by comparing the OCZ for sequences
06SL1e3 and 07SL1e3.  For effective temperatures below $\approx 22000$
K, the base of the OCZ is slightly shallower for the more massive
model. This is a consequence of the fact that the stellar envelope is
substantially denser for massive models.  As a result, conduction
becomes the main mechanism to transport energy in those regions, thus
inhibiting the occurrence of convection.  When we come to compare the
evolution of the OCZ for white dwarf models with the same stellar mass
and identical helium content, but with a different chemical structure
of their envelopes, the results depend on the mass of the helium
layer.  In the case of thick helium envelopes, the opacity is larger,
since double-layered models have more dredged carbon in their outer
layers, and the OCZ is in turn shallower for $T_{\rm eff} \lesssim
12500$ K as compared to the case of its single-layered counterpart.
In the case of the thin envelope, the OCZ is identical for the
chemical structure of both single- and double-layered models.  This is
expected because the thin helium envelope for single-layered models is
contaminated with some carbon, and thus the situation is similar to
what happens in the case of the double-layered models.

\subsubsection{Evolution of the chemical profiles}

\begin{figure}[ht]
\centering
\includegraphics[clip,width=250pt]{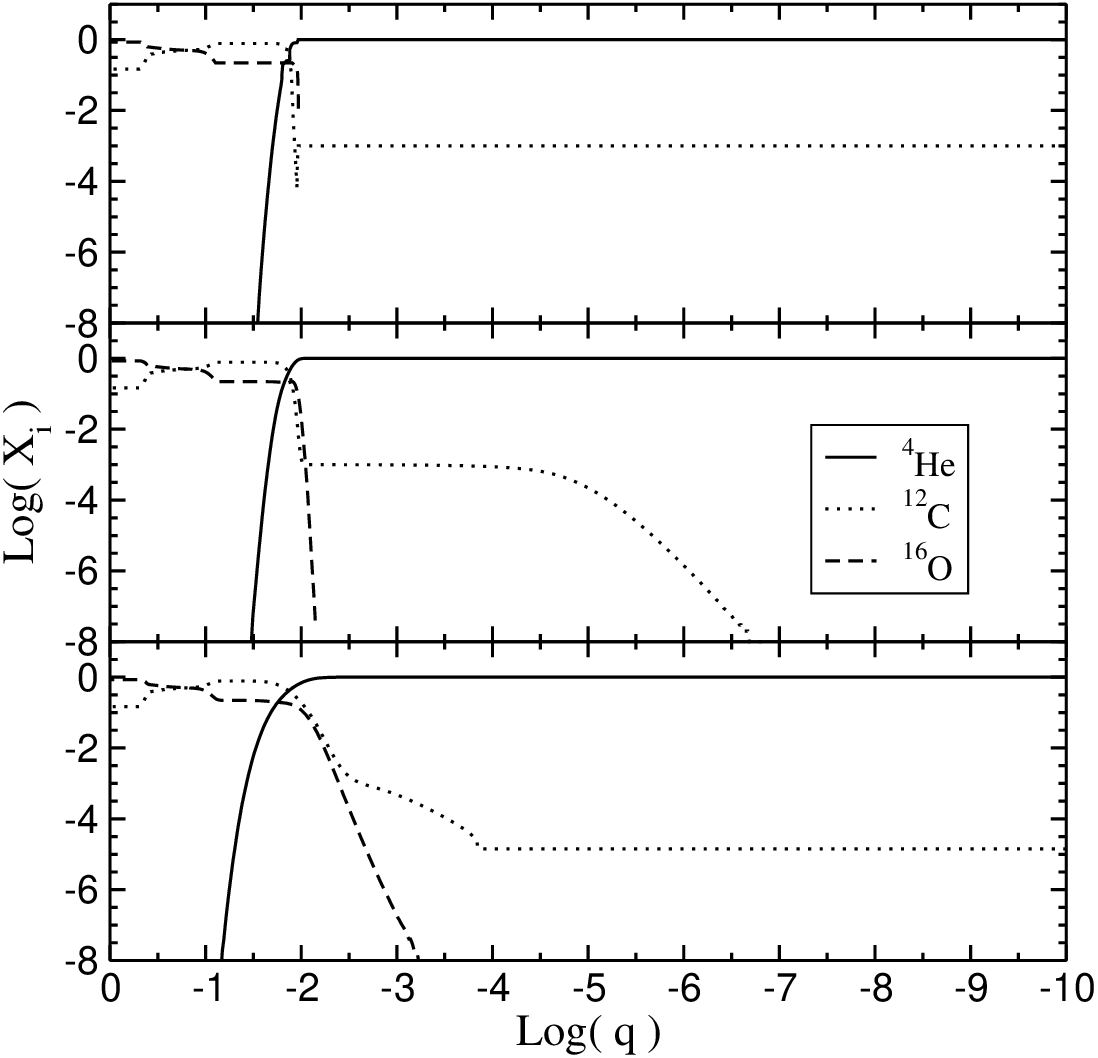}
\caption{ Abundance (by mass) of $^{4}$He, $^{12}$C, and 
$^{16}$O  in terms  of  the outer  mass  fraction for  0.5-$M_{\odot}$
single-layered  models with  $M_{\rm He  }= 10^{-2}  M_*$.   The upper
panel shows the situation at the beginning of the sequence. The middle panel  
depicts the  profiles at  $T_{\rm eff}=
25700$ K, and  the bottom panel displays the predictions  at $T_{\rm eff}=
7300$ K.  Note the presence of carbon in the outer layers as a result
of convective dredge-up.}
\label{perfilSL}
\end{figure}

As we shall see, the shape of the initial carbon distribution in the
intershell region determines how much carbon will eventually be
dredged up to the outer layers.  We begin by examining Fig.
\ref{perfilSL}, in which the abundances of $^{4}$He, $^{12}$C, and
$^{16}$O are plotted in terms of the outer mass fraction $q$ [$\equiv
(1-M_r/M_*)$] for 0.5-$M_{\odot}$ white-dwarf models with $M_{\rm He
}= 10^{-2} M_*$ and a single-layered envelope.  The figure documents
the situation at three different effective temperatures.  The top
panel corresponds to the initial model of the sequence 05SL1e2. The
envelope is made of a mixture of helium plus carbon traces ($X_C$=
0.001 by mass).  The action of diffusive processes even in deeper
layers is already evident in the middle panel, at $T_{\rm eff}= 25700$
K.  Note that at this stage, carbon has begun to sink into deeper
layers due to gravitational settling, whereas at the base of the
envelope the helium profile has been smoothed out by chemical
diffusion, leading to a non-zero helium abundance in layers originally
void of helium.  The bottom panel documents the situation at $T_{\rm
eff}= 7300$ K.  At low effective temperatures, the OCZ has increased
substantially and its base has reached the diffusive tail of the
carbon distribution.  As a result, considerable amounts of carbon have
been dredged up to the surface. Clearly, the final carbon abundance at
the surface of the star will be strongly dependent on the shape and
extent of the internal carbon profile\footnote{We remind the reader
that we have assumed carbon to be fully ionized in our treatment of
diffusion. This simplification may lead to an overestimation of the
surface carbon abundances at such very low \teff\ (see MacDonald et
al. 1998).}.

\begin{figure}[ht]
\centering
\includegraphics[clip,width=250pt]{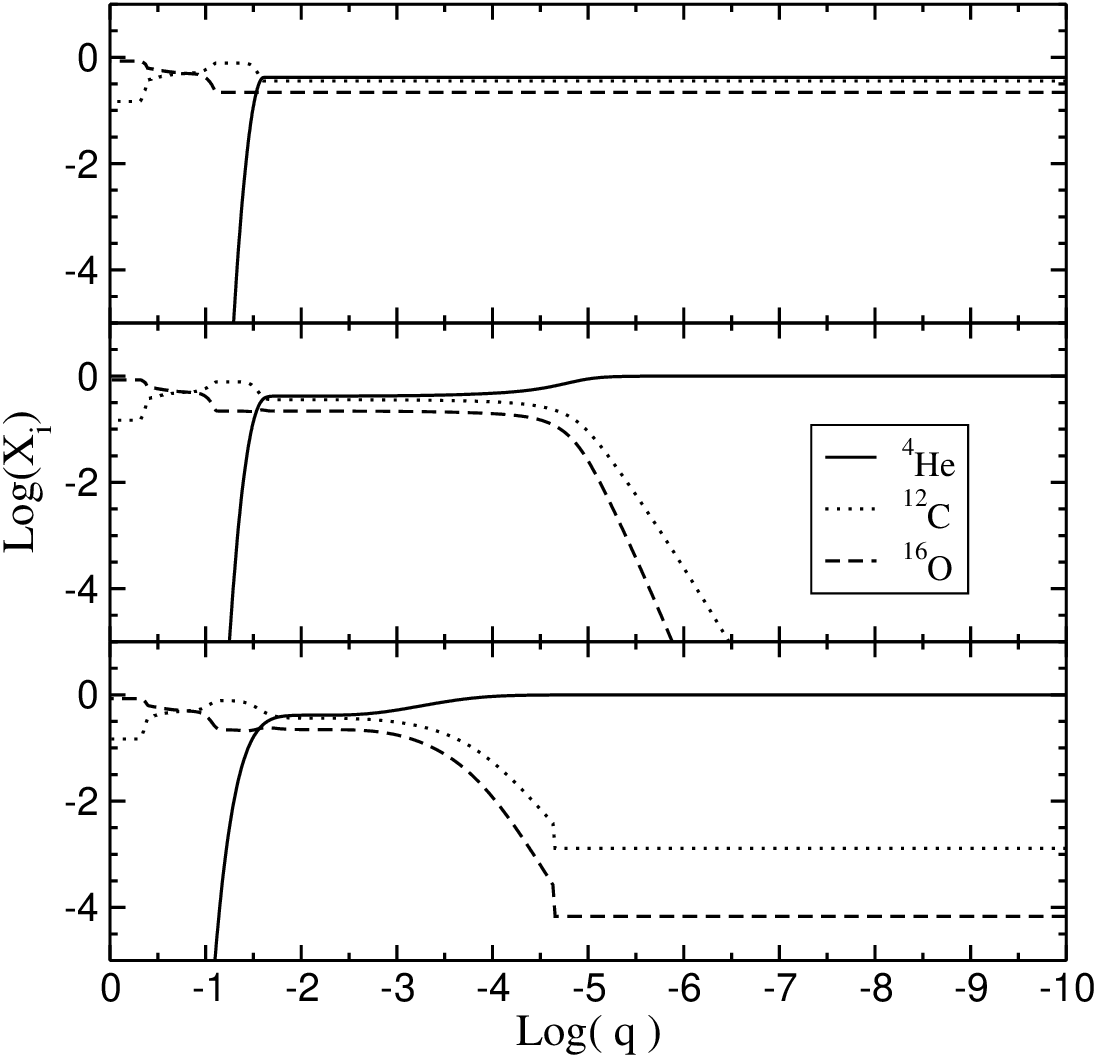}
\caption{Same as Fig. \ref{perfilSL}, but for double-layered models. 
Upper, middle, and bottom panels depict the situation at 49700, 22800,
and 8900 K, respectively.
}
\label{perfilDL}
\end{figure}

The situation for the corresponding 05DL1e2 double-layered sequence is
shown in Fig. \ref{perfilDL}. Again, the upper panel corresponds to
the initial model of the sequence, before diffusion is allowed to
operate. In this case, the envelope consists of a mixture of helium,
carbon, and oxygen, with abundances of 0.42, 0.36, and 0.22 by mass,
respectively. The overwhelming initial oxygen abundance is due to the
occurrence of overshoot episodes during the AGB and post-AGB evolution
of the progenitor star (Herwig et al.1999). The middle panel depicts
the situation at $T_{\rm eff}= 22800$ K. It is apparent from this plot
that carbon and oxygen, abundantly present in the initial outer layer
chemical profile, have migrated to deeper layers as an effect of
gravitational settling.  Clearly, the star has developed a
diffusion-induced double-layered structure characterized by a pure
helium mantle, which thickens as cooling proceeds, overlying a still
uniform intershell region rich in helium, carbon, and oxygen, the
relics of the last helium thermal pulse. By the time the domain of the
DQ white dwarfs is reached, the thickness of the pure helium mantle
amounts to 0.0002 $M_*$. The bottom panel displays the stage in which,
at low effective temperatures ($T_{\rm eff}= 8900$ K), carbon and
oxygen have been dredged to outer layers.  At this point, the star
should exhibit significant amounts of $^{12}$C and $^{16}$O in its
spectra, in addition to strong features of helium.

\subsubsection{Evolution of surface chemical abundances}

\begin{figure}[ht]
\centering
\includegraphics[clip,width=250pt]{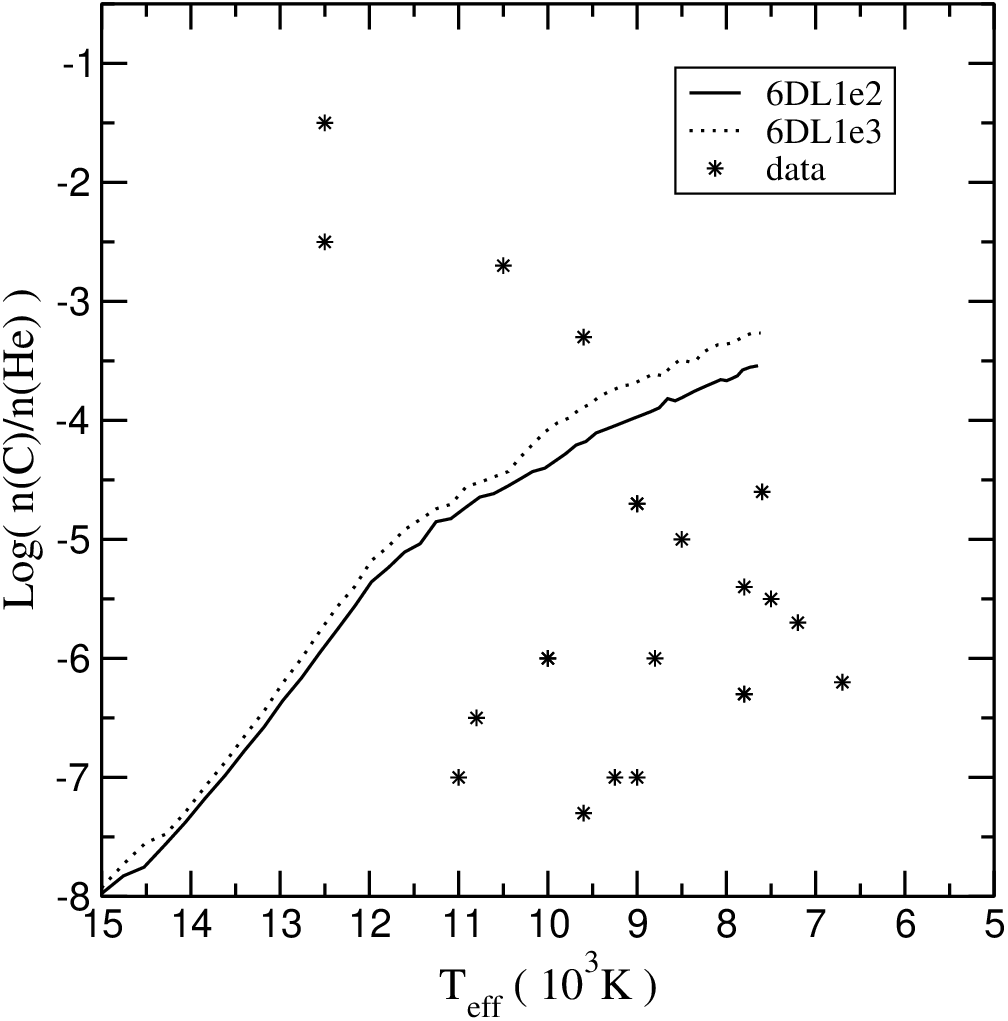}
\caption{Evolution of the surface carbon abundance 
(relative  to  helium  abundance) for  0.6-$M_{\odot}$  double-layered
models with $M_{\rm He }= 10^{-3} M_*$ (dotted line) and $M_{\rm He }=
10^{-2} M_*$ (solid line).  Dots represent the observational situation
as quoted by Weidemann \&  Koester (1995).}
\label{envolDL}
\end{figure}

In order to roughly estimate the possible observational consequences
 of having different {\it initial} structures for the outer envelopes
 (i.e. single- and double-layered envelopes), we now examine the
 evolution of the surface carbon abundance in terms of $\log(n_{\rm
 C}/n_{\rm He})$, with $n_i$ the surface number density of the species
 $i$. We expect the surface carbon abundance to grow as the DQ white
 dwarfs cool, because more and more carbon should be dredged up to the
 surface when the OCZ gets deeper. In the single-layered picture,
 carbon appreciably diffuses from the core to eventually meet the base
 of the helium convection zone, whilst for the double-layered models,
 the carbon pollution occurs when the convection zone reaches the tail
 of the {\it settling} carbon. The number density of surface $^{12}$C
 relative to $^{4}$He is shown in Fig.  \ref{envolDL}, for sequences
 06DL1e2 and 06DL1e3.  Observed photospheric ratios given by Weidemann
 \& Koester (1995) are also shown in the figure for comparison.  For
 the same stellar mass and initial outer layer chemical stratification
 (in this case a PG1159-like profile), we want to see what (if any)
 the effect of varying the content of helium is.  The figure indicates
 that the carbon abundances are very similar throughout the entire
 interval of $T_{\rm eff}$, where $\log(n_{\rm C}/n_{\rm He})$ is only
 marginally larger for the thin envelope models.  This is due to the
 fact that for thinner helium envelopes, carbon will reach more
 external layers than in the case of thicker envelopes. Note that
 double-layered models reproduce neither the DQs with low carbon
 abundances nor those with very high carbon abundances.

\begin{figure}[ht]
\centering
\includegraphics[clip,width=250pt]{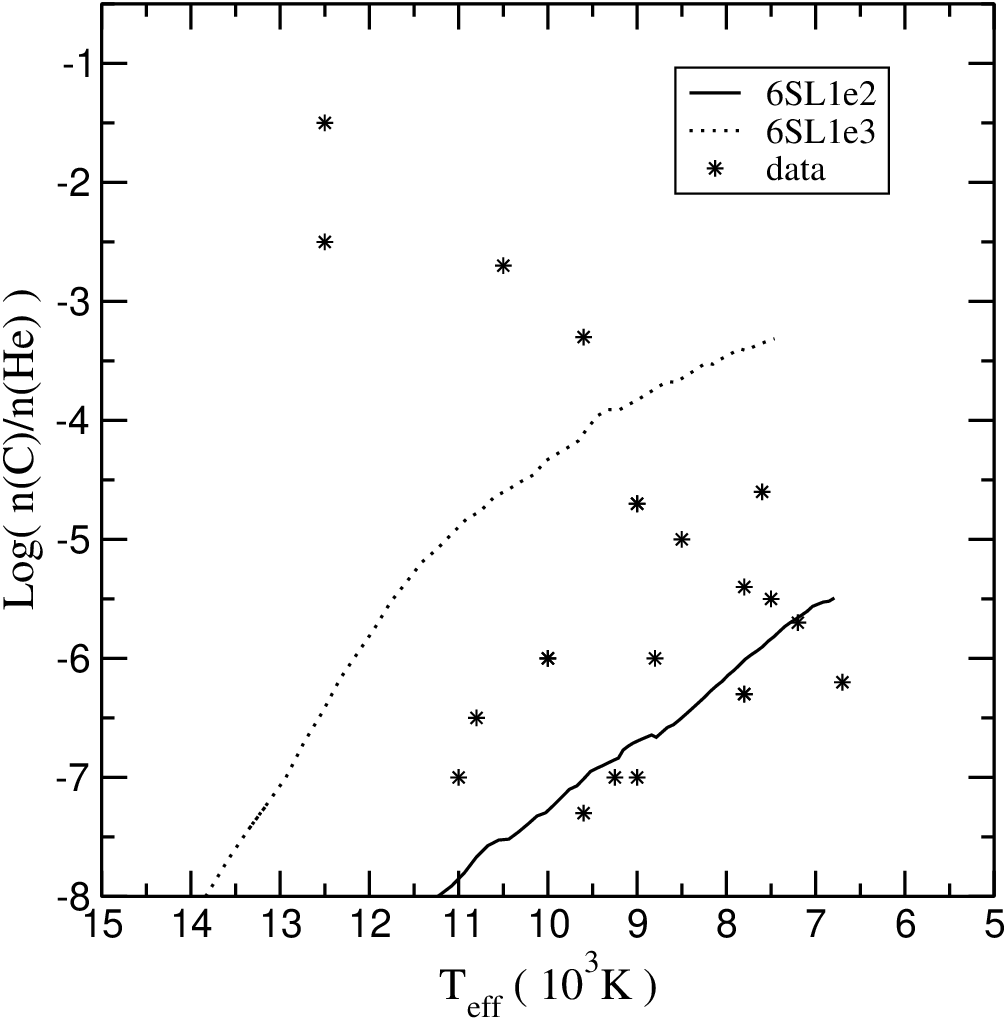}
\caption{Same as Fig. \ref{envolDL} but for single-layered models. 
Remarkably,  thin  envelope  models  exhibit  a  rather  large  carbon
abundance as compared with the case of thick envelope models. See text
for additional details.}
\label{envolSL}
\end{figure}

 The situation becomes quite different if single-layered models are
assumed from the start of the evolution. Indeed, note in
Fig. \ref{envolSL} that the predicted photospheric abundance of carbon
is markedly distinct when single-layered models with different helium
contents are taken into account.  The figure clearly shows that
thinner envelope models have a remarkably larger carbon abundances
than that of thicker envelope models.  Note that for models with
$M_{\rm He}= 10^{-3} M_*$ the carbon abundance is comparable to what
is predicted by their double-layered counterpart (compare with
Fig. \ref{envolDL}), whereas for models with $M_{\rm He}= 10^{-2} M_*$
the abundance of carbon is exceedingly low and in agreement with the
lowest detected carbon abundances. The behaviour displayed in
Fig. \ref{envolSL} can be explained on the basis that the envelope
initially consists mostly of pure helium in a single-layered model.
If the envelope is thick enough, the tail of the carbon distribution
is located at such a depth in the star that the amount of carbon that
is eventually dredged up by convection is much less than in the case
of a thin envelope.

\begin{figure}[ht]
\centering
\includegraphics[clip,width=250pt]{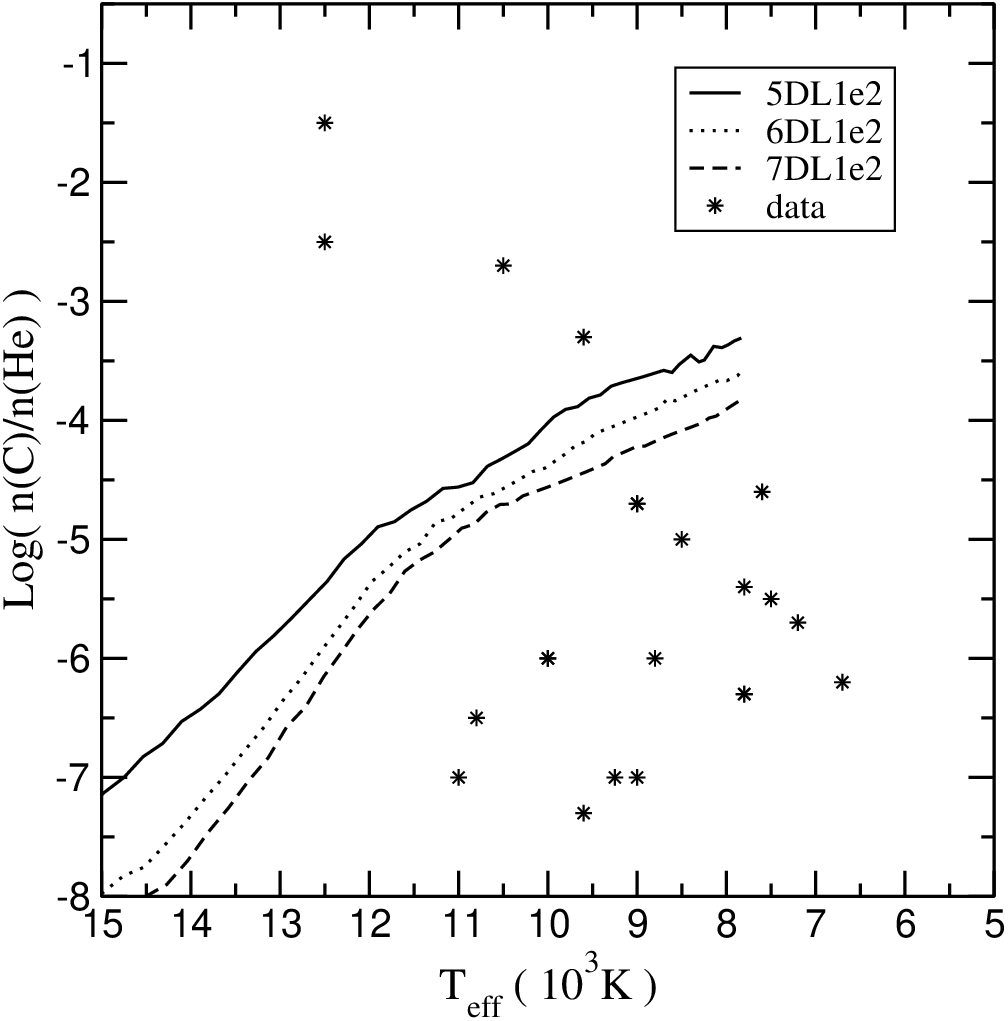}
\caption{Same as Fig. \ref{envolDL} but for  double-layered models
with different stellar masses and equal helium content.}
\label{masseffect}
\end{figure}

We now turn to the question of whether the stellar mass has any
influence on the carbon abundance expected in DQ white dwarfs. In
Fig. \ref{masseffect} we show the number density of surface $^{12}$C
relative to $^{4}$He as a function of the effective temperature for
three sequences of double-layered models with the same helium content
$M_{\rm He}= 10^{-2} M_*$ and stellar masses of 0.5, 0.6, and 0.7
$M_{\odot}$. We note that the carbon abundance decreases as the
stellar mass increases. This behaviour is the one expected, since in
more massive white dwarfs conductive transport of energy prevents the
OCZ from growing into deeper layers.

\begin{figure}[ht]
\centering
\includegraphics[clip,width=250pt]{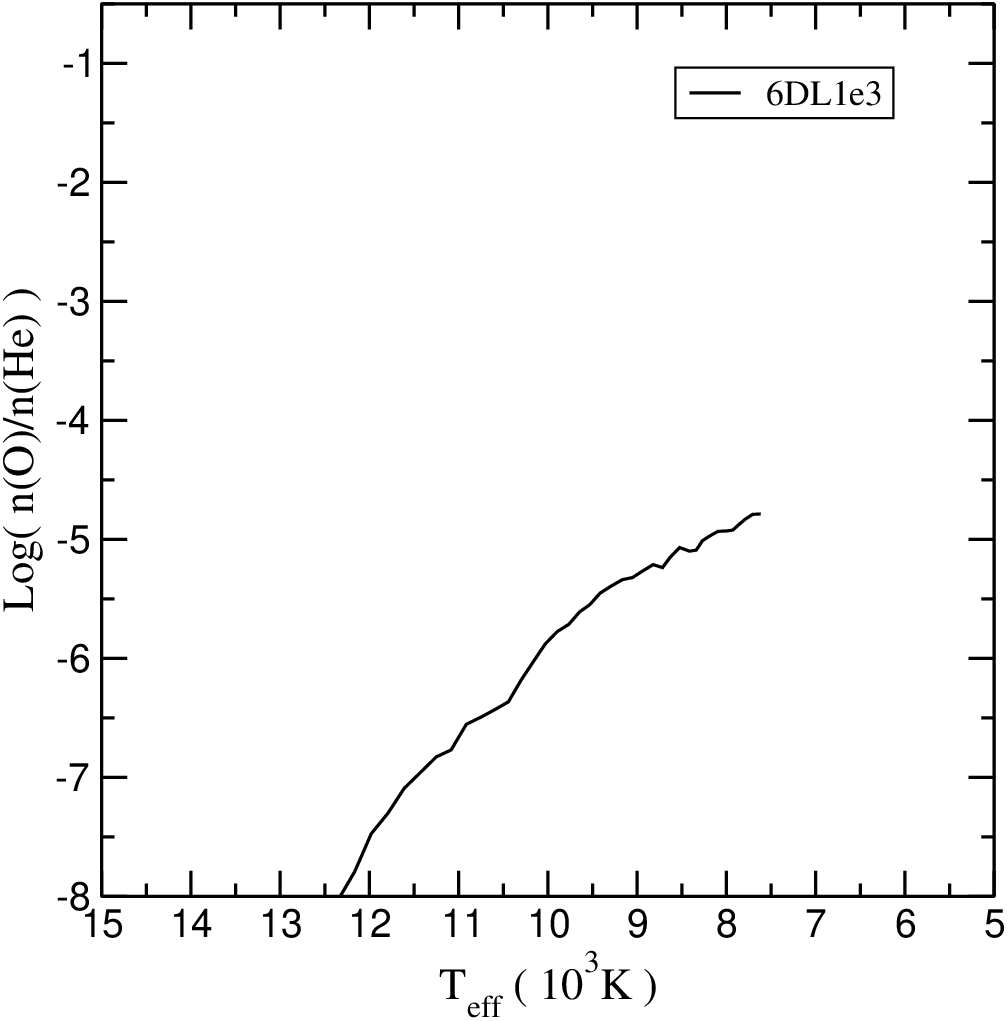}
\caption{The evolution of the surface oxygen abundance 
(relative  to helium) for  0.6-$M_{\odot}$ models  with $M_{\rm  He }=
10^{-3} M_*$ and a double-layered chemical structure.}
\label{oxigeno}
\end{figure}

We have found that, at low effective temperatures, an appreciable
amount of oxygen should emerge at the surface of those DQ white dwarfs whose
initial envelope configurations are characterized by a double-layered
structure with high oxygen content. This expectation is borne out by
Fig. \ref{oxigeno}, which shows the evolution of the surface oxygen
abundance (in terms of the helium one) for a 0.6-$M_{\odot}$ model
with $M_{\rm He }= 10^{-3} M_*$ and a double-layered chemical
structure.

\subsubsection{The role of overshooting}

 In order to  explore the  possibility that  convective overshoot
could alter the expected surface carbon abundances, we considered
the occurrence of overshoot mixing  below the outer convection zone by
employing  the   simple  formalism  described  in   Maeder  \&  Meynet
(1989). In  particular, we  adopted the distance  of overshooting
$d_{\rm ov}$ as a  fraction of the pressure scaleheight $H_{\rm p}$
at the canonical border of the convective zone: $d_{\rm ov}=
\alpha_{\rm ov} H_{\rm p}$. We chose the values of 0.1, 0.2, and 0.7 
for $\alpha_{\rm ov}$, and calculated sequences with overshooting for
both DL and SL and different stellar masses.  For all the cases, we
find that, as expected, when overshoot episodes are taken into
account, the carbon abundance dredged to the surface increases, but
only slightly. Indeed, the carbon abundance increases by at most 0.48
dex when a large overshooting is adopted. Needless to say, if
overshooting takes place in these stars, a link between DQs with low
carbon abundances and PG1159 stars turns out to be even less probable.

\subsection{The AGB-manqu\'e star}

In what follows, we consider the evolution of a 0.49-\msun\ star model
representative of an AGB manqu\'e star.  As mentioned in the
introduction, these objects are carbon-normal stars characterized by
a very thin hydrogen envelope and evolving directly from the EHB into
the white-dwarf state by eluding the AGB stage.  For this sequence we
 consider a hydrogen-deficient, helium-rich, carbon-trace
($X_C$= 0.03) initial stellar configuration. This model was 
obtained from a 1-\msun\ main sequence star.  We were specially
careful in the choice of the initial surface carbon abundance in
order to be consistent with observational data as well as with current
evolutionary calculations that lead to this type of EHB star.
Cassisi et al. (2003) and Lanz et al. (2004) have obtained models of
EHB stars by assuming that the core helium flash in low mass stars occurs
after the tip of the red giant branch, as a consequence of extreme
mass loss episodes. If this flash occurs along the white dwarf cooling
regime, the convection zone reaches the hydrogen-rich envelope, mixing
hydrogen into the hot helium-burning environment where protons are
burned rapidly.  As a result, the outermost layers become
hydrogen-poor and rich in helium, carbon, and oxygen.
\begin{figure}[ht]
\centering
\includegraphics[clip,width=250pt]{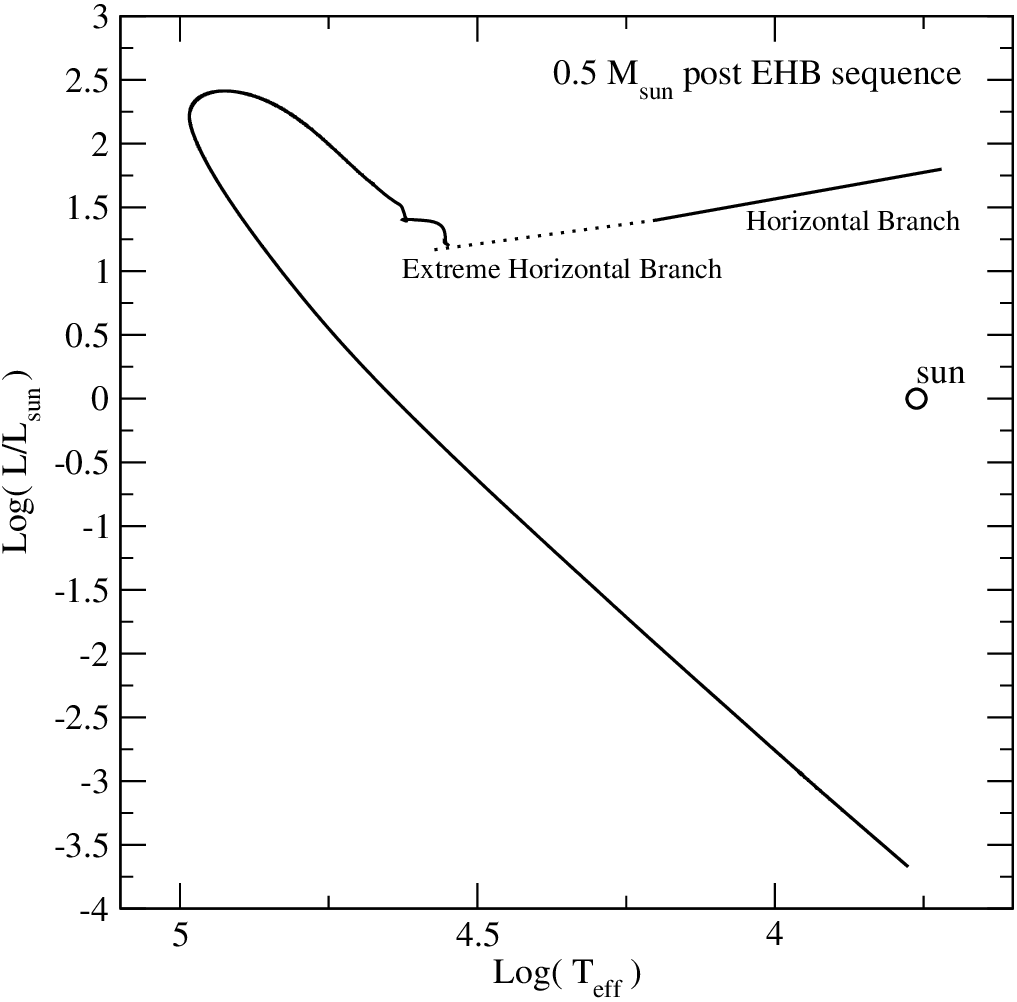}
\caption{ Evolutionary track for the AGB manqu\'e star. As a reference,
  the position of the sun is shown with the symbol $\circ$, as well
 as the location of the horizontal branch. The maximum effective
 temperature reached is 96500 K and the highest luminosity is
 log(L/L$_{\odot}$) = 2.4113.}
\label{hr_AGBmanque}
\end{figure}

\begin{figure}[ht]
\centering
\includegraphics[clip,width=250pt]{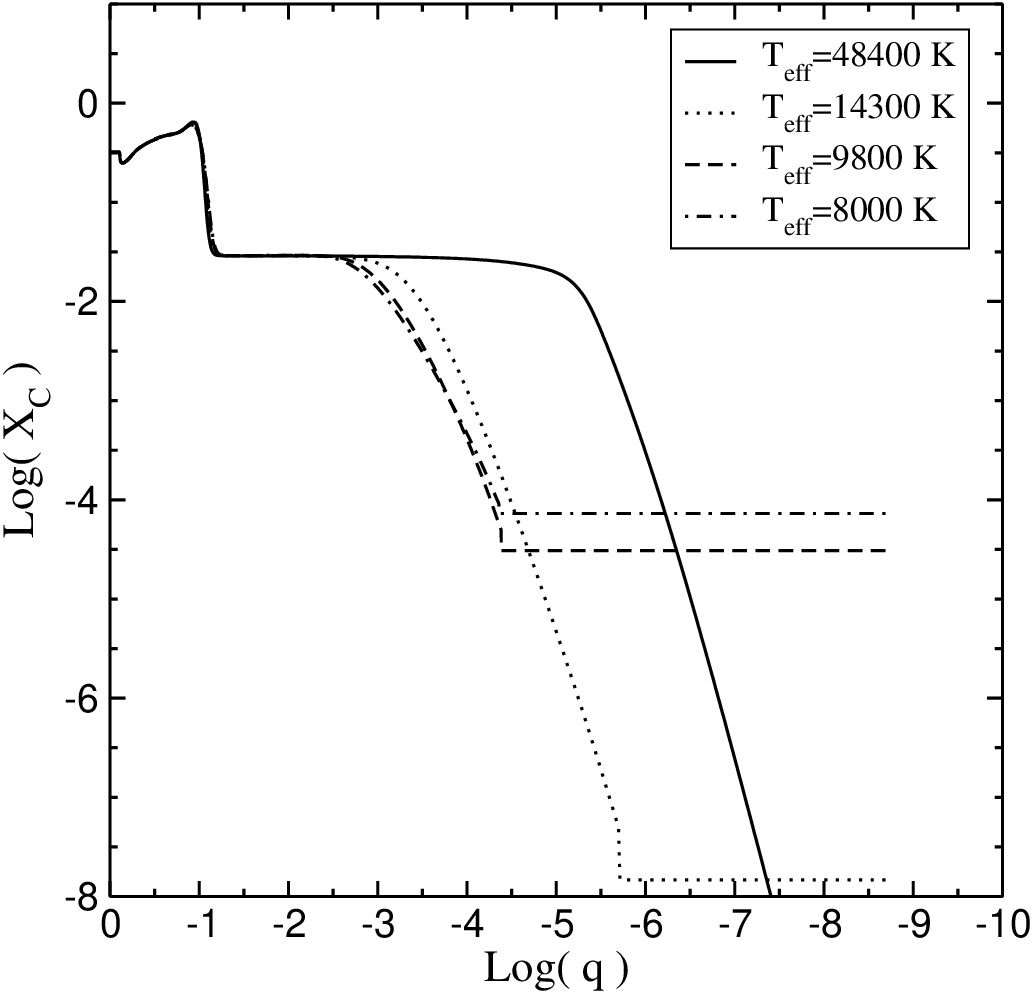}
\caption{Abundance (by mass) of $^{12}$C in terms of the 
outer mass  fraction for the 0.49-$M_{\odot}$  AGB-manqu\'e model star
at four different stages of the white-dwarf cooling.}
\label{perfilesAGB}
\end{figure}

\begin{figure}[ht]
\centering
\includegraphics[clip,width=250pt]{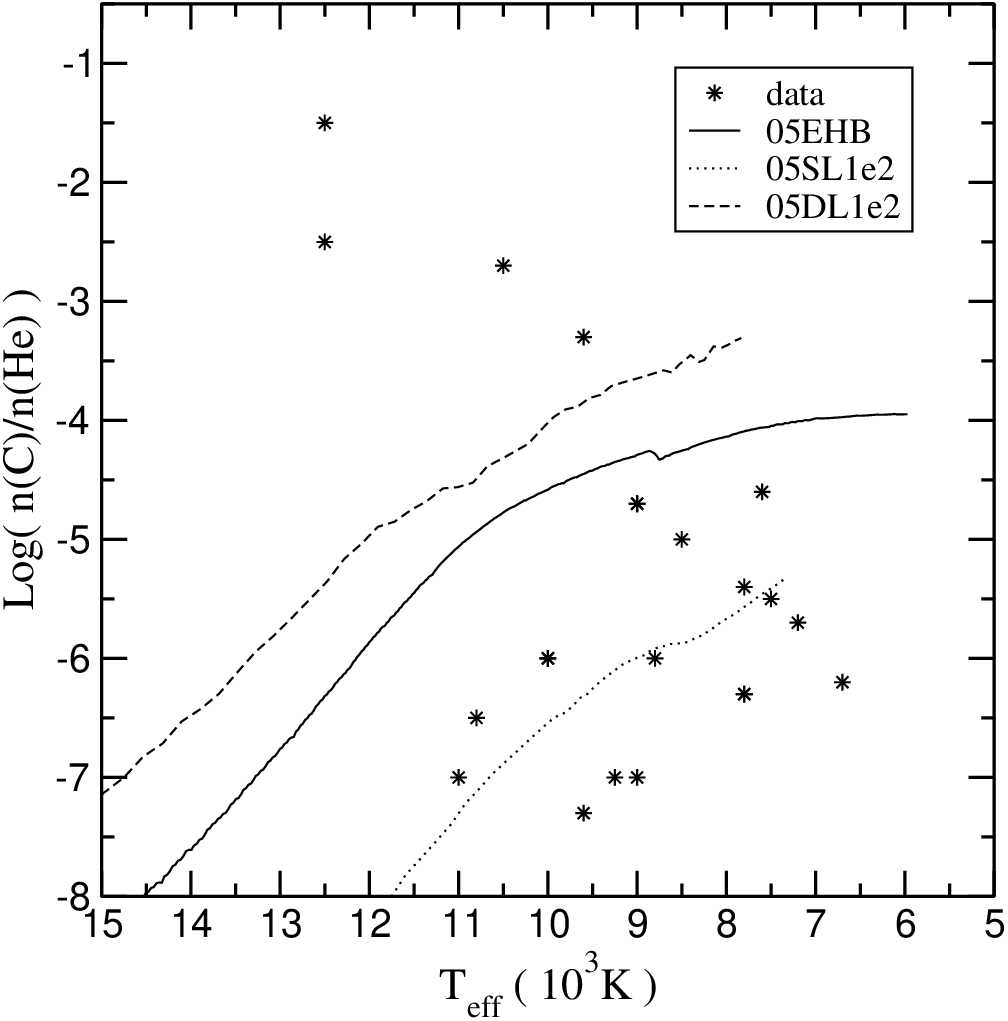}
\caption{The surface carbon abundance (relative to helium) 
for  the 0.49-\msun  AGB-manqu\'e  model. We  have  also included  the
carbon  abundances  for  the   sequences  05SL1e2  and  05DL1e2.  Also
displayed is  the observational situation,  as quoted by  Weidemann \&
Koester (1995).}
\label{abunAGBmanque}
\end{figure}

We followed the evolution of the 0.49-\msun\ star from the EHB
throughout the white-dwarf stage to the DQ regime (sequence 05EHB).
In Fig. \ref{hr_AGBmanque} we show its evolutionary track. Note that
the star evolves directly from the horizontal branch to the
white-dwarf cooling track. After reaching a maximum effective
temperature value, the remmant star enters its final white-dwarf
cooling branch. The evolution of the carbon chemical profile is shown
in Fig.  \ref{perfilesAGB}, where we depict the carbon chemical
profile at four selected stages.  When the DQ regime is reached, the
stellar models present a surface carbon abundance that is smaller than
in the double-layered models, but larger than that of the
single-layered models, as can be seen in Fig. \ref{abunAGBmanque}. The
latter trend is due to the large initial carbon abundance as compared
to the one assumed in single-layered models.

It is important to note that there are observational examples of EHB
stars with very low carbon abundances\footnote{For example, Lanz et
al.  (2004) obtained a carbon abundance (by mass)  for LB1766 of
$10^{-4}$.} that cannot be explained by mixing events due to a helium
flash or by any current evolutionary scenario. From our preceding
results for single-layered models, it is expected that these
carbon-poor post-EHB stars could evolve into DQ white dwarfs with 
{\it very low} surface carbon abundances.

\section{Implications of our results}

The DQ stars are evolved helium-rich white dwarfs characterized by the
presence of abundant carbon in their outer layers. The origin of
carbon in the atmospheres of such white dwarfs has been interpreted as
the result of the dredge-up of the diffusive tail of carbon by the
superficial helium convection zone. These stars play an important role
in the quest for interpreting the spectral evolution in white dwarfs.
An interesting feature of these stars is that they exhibit a
wide range of surface carbon abundances, which could be reflecting a
marked diversity in the evolutionary history of the progenitor stars.
It is in this context that we undertook this
investigation, where our main aim was to explore the implications of
the different formation channels for chemical stratifications in the
outer layers of DQs.  In particular, we tried to establish a
connection between the DQs characterized by low carbon abundances and
the different evolutionary channels through which hydrogen-deficient
white dwarfs could be formed.

Theoretical and observational evidence exists that fosters the
plausibility of an evolutionary connection between most of the
post-born-again hydrogen-deficient PG1159 stars with the evolved DQ
white dwarfs via the link PG1159 -- DB -- DQ.  Our results strongly
suggest that, if the canonical convective dredge-up is the source of
observed carbon, those DQ white dwarfs with very low detected carbon
cannot be linked to the PG1159 stars, results which reinforce the
conclusions arrived at in Althaus \& C\'orsico (2004).  We have found
 that this conclusion remains valid irrespective of the stellar
mass and helium content with which the PG1159 stars are formed. This
prompts us to suggest that a fraction of the observed DQ population
could have followed an alternative evolutionary path that could link
them with stars that have somehow avoided the thermally pulsing AGB
phase. In addition, we find that PG1159 also does not constitute a viable
route to explain those DQs with very high carbon abundances.

 In view of these concerns, we investigated alternative scenarios for
the formation of hydrogen-deficient white dwarfs. The hydrogen-poor
EHB stars and the supergiant, helium-rich RCrB stars constitute an
attractive possibility. The EHB stars are particularly interesting,
because after exhausting helium in their cores, they evolve directly
to the white dwarf state.  That these stars do not experience the
recurrent AGB thermal instabilities is expected to lead to white dwarf
progenitors with an outer layer chemical stratification that is quite
different from what is expected for a post-born-again PG1159 star.
Indeed, their envelopes resemble one for a single-layered
profile. Although the evolution of the AGB manqu\'e star presented
here shows a rather high C-to-He ratio, it is important to notice that
the initial C abundance is compatible with evolutionary calculations
for the formation of these stars. However, stars do exist in the EHB
with very low carbon abundance, which would lead to lower C to He
ratios.  As our results in connection with single-layered models with
thick helium envelopes suggest, it is not inconceivable that these
stars could be the ancestors of the DQ population characterized by low
surface carbon abundance. Also the extreme helium stars (and the
associated RCrB stars) evolve directly to the white dwarf
regime. These stars have single-layered chemical structure and could
evolve into DQ white dwarfs with very low carbon abundances, provided
their initial carbon abundance is $\sim$ 0.001. An observational
counterpart of this expectation is provided by MV Sgr, a hot
helium-rich RCrB star characterized by a very low surface carbon
abundance (see Table 5 in De Marco et al. 2002).

In closing, we stress again that our conclusions are based on the
assumption that the canonical convective dredge-up is the mechanism
responsible for the observed carbon abundances in DQ white
dwarfs. However, additional mechanisms could be operating at least in
hotter helium-rich white dwarfs.  As a matter of fact, there is ample
evidence that the atmospheres of hot DB white dwarfs are polluted by
traces of carbon, which cannot be explained solely in terms of
convective dredge-up. For instance, Provencal et al. (2000) present
evidence in favour of horizontal motions as a viable mechanism by
which inner carbon could be transported to the stellar surface in some
hot DBs. Finally, we want to mention that on the basis of
chemical homogeneous atmosphere models, we find that a moderate
presence of carbon in the outer, radiative layers could hinder the
optical detection of high carbon abundances in the convective
region. These preliminary results suggest that the relatively low
carbon abundances frequently observed in cool He-rich WDs may be
explained by a carbon-stratified photosphere, where carbon-rich
convective layers remain partially hidden by an enhanced He$^-$
opacity in superficial layers ($\tau_{Ross} < 1$). We postpone
the discussion of this topic to a further investigation.

\begin{acknowledgements}
We acknowledge the comments and suggestions of an anonymous referee
that greatly improved the original version of this paper.  This
research was supported by the Instituto de Astrof\'{\i}sica La
Plata. A. M. S. was supported by the National Science Foundation
through the grant PHY-0070928.

\end{acknowledgements}

\end{document}